\documentclass{article}

\usepackage{spconf, graphicx, amsmath, hyperref,setspace,enumitem, amssymb, bm}
\usepackage{subcaption}
\usepackage{multirow}
\usepackage[dvipsnames]{xcolor}
\usepackage{mathtools}
\usepackage[ruled]{algorithm2e}
\usepackage{hyperref}

\title{Generative De-Quantization for Neural Speech Codec\\via Latent Diffusion}
\name{Haici Yang$^1$, Inseon Jang$^2$, and Minje Kim$^1$\thanks{This work was supported by Electronics and Telecommunications Research Institute (ETRI) grant funded by the Korean government
(23ZH1200; ``The research of the basic media contents technologies").}}
\address{$^1$Indiana University, Department of Intelligent Systems Engineering, Bloomington, IN, USA 47408\\
$^2$Electronics and Telecommunications Research Institute, Daejeon, South Korea 34129}

\begin{document}
\ninept
\maketitle
\begin{abstract}



In low-bitrate speech coding, end-to-end speech coding networks aim to learn compact yet expressive features and a powerful decoder in a single network. A challenging problem as such results in unwelcome complexity increase and inferior speech quality. In this paper, we propose to separate the representation learning and information reconstruction tasks. 
We leverage an end-to-end codec for learning low-dimensional discrete tokens and employ a latent diffusion model to de-quantize coded features into a high-dimensional continuous space, relieving the decoder's burden of de-quantizing and upsampling. To mitigate the issue of over-smooth generation, we introduce midway-infilling with less noise reduction and stronger conditioning. In ablation studies, we investigate the hyperparameters for midway-infilling and latent diffusion space with different dimensions. Subjective listening tests show that our model outperforms the state-of-the-art at two low bitrates, 1.5 and 3 kbps. 
Codes and samples of this work are available on our webpage 
\footnote{\url{https://saige.sice.indiana.edu/research-projects/LaDiffCodec}}. 

\end{abstract}

\section{Introduction}
\label{sec:intro}

Neural speech codecs are designed to capture human speech's intricate patterns more effectively than traditional methods. 
Recently, with the successful attempts of codec-based generation \cite{borsos2023audiolm, Wang2023synth, kreuk2022audiogen}, high-bitrate neural codecs \cite{Zeghidour2021soundstream, Kumar2023arxiv, defossez2022highfi} gain much attention, for its capability of recovering high-fidelity audio. 

Current high-fidelity codecs mostly perform waveform coding \cite{ZhenK2020spl, PetermannD2021harpnet}. They typically learn encoder and decoder models end-to-end, aiming at extracting expressive latent features and powerful decoders in a single network. Reconstruction objectives favor preserving essential information. Thus, with abundant data, end-to-end models are generally good at learning representations and achieving high-fidelity reconstruction at higher bitrates.
For example, SoundStream achieves reasonable speech quality at 3kbps with its fully convolutional architecture and residual vector quantization \cite{Zeghidour2021soundstream}. Pre-trained transformer and language models have been used to assist low-bitrate coding in open-sourced EnCodec \cite{defossez2022highfi} and an ultra-low-bitrate codec \cite{SiahkoohiA2022interspeech}. More recently, DAC \cite{Kumar2023arxiv} implements RVQGAN and snake activation into the architecture and improves coding quality, especially on high bandwidth. Many other improvements involve disentanglement \cite{YangH2021sanac, jiang2023disentangled, omran2023disentangle} into end-to-end coding to remove redundancy.
However, the above works achieve good quality speech only at medium or high bitrate ($\ge$ 3 kbps).  In the low-bitrate case, excessively deep and complex networks are necessary to learn low-dimensional representation, which can impair end-to-end training. Learning expressive features and powerful decodes at the same time becomes challenging. 

Therefore, on coding low bitrate, many works turn to a vocoder-based codec to leverage generative models, and reconstruction of speech from existing features, e.g., WaveNet-based codecs \cite{KleijnW2018wavenet,GarbaceaC2019vqvae}, LPCNet for real-time coding \cite{ValinJ2019lpcnetcoding}, GAN-empowered resynthesis \cite{polyak2021speech}.
 Recent employment of AudioLM \cite{borsos2023audiolm} in SoundStream as in LMCodec \cite{JenrungrotT2023lmcodec}, reduces the bitrate down to $\sim$ 1kbps. Neural feature predictor \cite{YangH2023lpcnet} applies a generative model on the feature domain to assist LPCNet with more efficient feature input. 
 
\begin{figure}
    \centering
    \includegraphics[width=\columnwidth]{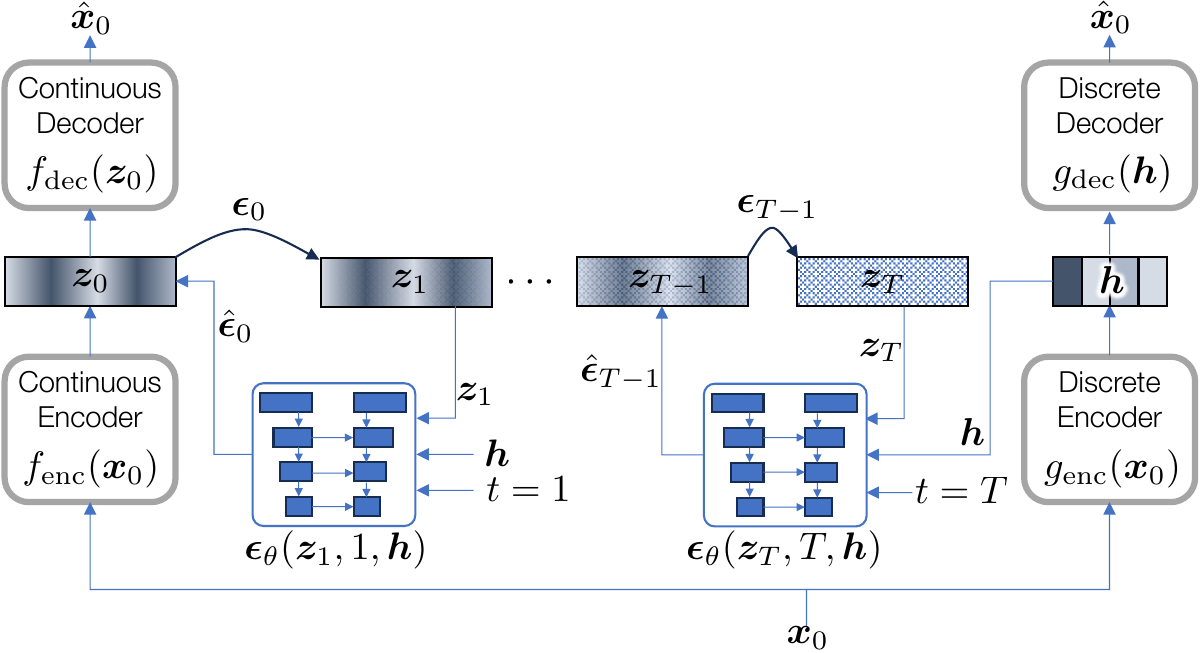}
    \caption{A diffusion model de-quantizes the low-dimensional, low-bitrate discrete speech tokens into high-dimensional continuous variables. }
    \label{fig:overview}
\end{figure}
We observed from pilot experiments that low-bitrate coded features from waveform codec can preserve distinguishable speech features and better capture essential information than the traditional speech features at the same bitrate. Thus, it is worth utilizing the representation learned from the end-to-end audio coding schemes while replacing the end-to-end trained decoders with more powerful generative models. 
To this end, the proposed system consists of three modules. First, we employ an autoencoder whose bottleneck is defined by a continuous and high-dimensional latent space. We assume that its decoder is responsible for high-fidelity reconstruction as it is exempt from the de-quantization and dimension expansion tasks. Second, we employ another end-to-end codec that properly performs the dimension reduction and quantization task in its bottleneck. This end-to-end codec will perform coarse reconstruction from its low-bitrate code, but high fidelity is not guaranteed. Finally, we employ a latent diffusion model that bridges the gap between the two feature representations. By conditioning the diffusion model with the lower-dimensional quantized code from the end-to-end codec, we expect this model to perform generative de-quantization and upsampling tasks. 

We opt for the diffusion model among other generative models because 1) diffusion models have been successful in generating natural-sounding speech and audio \cite{pascual2023full}, from an extensive range of conditions;
2) Diffusion models provide a way to generate the entire feature map altogether and refine it iteratively. In contrast to the autoregressive models, diffusion models look at a large condition space, which significantly increases the quality upper bound. 
Specifically, we demonstrate that a latent diffusion (LD) model \cite{rombach2022high} is suitable in this problem setup, because running on the latent space offloads the diffusion model's potential burden to reconstruct raw waveforms: Instead, the LD model focuses on generating latent feature vectors. We explore the trade-off between quality and efficiency in choosing a diffusion space. As diffusion models are prone to generate over-smooth speech and hallucinate content, we introduce a new sampling technique that adds stronger prior to the conditional generation. We notice a concurrent work that explores high-fidelity audio generation from the speech codec by multi-band diffusion \cite{roman2023discrete}. In comparison, our model focuses on speech generation and achieves better quality at both low- and high-bitrate cases with only one \textit{latent} diffusion model, thus more efficient with respect to the model design. We call our model the latent diffusion codec (LaDiffCodec).


\section{Generative De-quantization with Latent Diffusion}

Fig. \ref{fig:overview} provides an overview of the proposed LaDiffCodec. It consists of the latent diffusion process in the middle, which converts the quantized code generated from the discrete coding module (on the right) into the continuous code (on the left). 

\subsection{Latent Diffusion}
Diffusion models are generative models characterized by two Markov processes: diffusion and reverse processes.
The diffusion process, $q(\bm x_{1:T}| \bm x_0) = \prod^T_{t=1}q(\bm x_t|\bm x_{t-1})$, corrupts clean data point $\bm{x}_0$ by gradually adding Gaussian noise, until it reaches a random variable $\bm x_T$ that is close to the standard normal distribution. Hence, $ q(\bm x_t| \bm x_{t-1}) \sim \mathcal{N}(\sqrt{1-\beta_t}\bm x_{t-1}, \beta_tI)$, where $\beta_t$ is the pre-defined noise schedule ($0 < \beta _0 < ... < \beta_T < 1$).
With reparameterization, sampling a step from the diffusion process can be acquired by, $\mathcal{F}: \bm x_0 \mapsto \bm x_t$,
\begin{equation}
    \mathcal{F}(\bm x_0, t)  = \sqrt{\bar{\alpha}_t}\bm x_0 + \sqrt{1-\bar{\alpha}_t} \bm \epsilon,
\end{equation}
where $\bm \epsilon \sim \mathcal{N}(0, 1)$ and $\bar{\alpha}_t = \prod_{i=0}^t(1-\beta_i)$. 

The learned reverse process $ p_\theta(\bm x_{0:T}) = p(\bm x_T)\prod^T_{t=1}p_\theta(\bm x_{t-1}|\bm x_{t})$ is commonly represented by a parametric function, e.g., a neural network, which predicts noise $\bm \epsilon_t$ at step $t$ to denoise the data point $\bm x_t$. This process can be conditioned by various types of auxiliary information.

While diffusion models succeed on many data generation tasks, the high computational complexity limits their accessibility. Meanwhile, the models are prone to spend excessive amounts of resources on modeling imperceptible details, especially on the multi-modality tasks \cite{rombach2022high}. As a solution, latent diffusion models propose to operate in the latent space $\bm z$ that is usually learned by a pre-trained autoencoder
\begin{equation}
    \hat{\bm{x}}_0 \leftarrow f_\text{dec}(\bm z_0), \quad \bm z_0 \leftarrow f_\text{enc}(\bm{x}_0),
\end{equation}
assuming this space to be computationally preferable and perceptually equivalent to the data domain.

For example, in LD, the diffusion process is defined in the latent space: $q(\bm z_{1:T}| \bm z_0) = \prod^T_{t=1}q(\bm z_t|\bm z_{t-1})$.

\subsection{Diffusion-Based De-Quantization }
LaDiffCodec maps the two latent spaces, the low-dimensional discrete code $\mathbb{H}$ and the high-dimensional continuous feature $\mathbb{Z}$. The restorative nature of this mapping requires conditional generation. In this section, we describe the three components of LaDiffCodec: discrete coding, continuous coding, and conditional diffusion sampling.

\noindent\textbf{Discrete coding}:
Our discrete coding module $g(\cdot)$ is an autoencoder-type codec that learns the discrete code space $\mathbb{H}$ using its encoder component, $g_\text{enc}: \mathbb{R}^N\rightarrow\mathbb{H}^D$. It provides the discretized features $\bm{h}\in\mathbb{H}^D$, which serve as the transmission bitstream in the ordinary codec usage. Ideally, we want the quantized speech tokens $\bm h$ to contain enough information for faithful speech reconstruction. Therefore we opt for state-of-the-art autoencoder codecs, which are designed to better capture coding features.  Existing codec relies solely on the decoder function $g_\text{dec}: \mathbb{H}^D \rightarrow \mathbb{R}^N$, which could be the performance bottleneck when $\mathbb{H}^D$ is low dimensional and discrete. In LaDiffCodec, we repurpose this discrete code $\bm h$ to condition the reverse diffusion sampling process. We employ Encodec \cite{defossez2022highfi} as the backbone of this discrete coding module.


\noindent\textbf{Continuous coding}:
To de-quantize the discrete token $\bm h$ into a continuous feature vector $\bm z$, LaDiffCodec utilizes an LD model defined in the continuous space $\mathbb{Z}$. 
To this end, we build another EnCodec-like continuous autoencoder, whose encoder maps the raw signal space $\mathbb{X}$ into the feature space $\mathbb{Z}$, i.e., $f_\text{enc}: \mathbb{X}^N\rightarrow\mathbb{Z}^L$, followed by a decoder that maps it back to the signal domain: $f_\text{dec}: \mathbb{Z}^L\rightarrow\mathbb{X}^N$. There is a trade-off in this continuous latent space. One may want the latent dimension $L$ to be as large as possible to keep its high expressiveness. However, with a diffusion model defined in this space, high dimensionality entails a long sampling time, causing reduced efficiency. Moreover, the gap between the high-dimensional continuous space and a low-dimensional discrete space has to be filled with upsampling layers, which brings additional artifacts. 
In our experiments, we investigate a few options of $L$ to minimize the tradeoff.

\noindent\textbf{Conditional Latent Diffusion}:
A diffusion model built on $\mathbb{Z}$ gradually adds noise $\bm\epsilon_t$ to $\bm z_t$ in its diffusion process. It employs a conditioned neural network model, which estimates $\bm\epsilon_t$ with the reweighted training objective for training the denoising (reverse) process \cite{ho2020denoising},
\begin{equation}
    \mathbb{E}_{\bm z_0, t, \bm h}(||\bm \epsilon_t - \bm \epsilon_{\theta}(\bm z_t, t, \bm h)||)
\end{equation}
where $\bm \epsilon_{\theta}(\bm z_t, t, \bm h)$ is parameterized by a neural network with weights $\theta$. It predicts the noise in the generation process. Hence, the loss is its expected difference from $\bm \epsilon_t$. 
The quantized features $\bm h$ condition both training and sampling stages to steer the generation. 

\subsection{Midway-Infilling}
The original sampling algorithm proposed in denoising diffusion probabilistic models (DDPM) \cite{ho2020denoising} iteratively removes predicted noise $\bm \epsilon_\theta$ from the noisy data samples,  $\mathcal{G} \colon \bm x_t \mapsto \bm x_{t-1}$, 
\begin{equation}\nonumber
    \mathcal{G}(\bm x_t, t, \bm h)\!=\!\frac{1}{\sqrt{1\!-\!\beta_t}}\!\left(\bm x_t \!-\! \frac{\beta_t}{\sqrt{1-\bar{\alpha}_t}} \bm \epsilon_\theta\big(\bm x_t, \sqrt{\bar{\alpha}_t}, \bm h\big)\right)+ \sqrt{\beta_t} \bm n,
\end{equation}
from $\bm x_T$ to $\bm x_0$, where $\bm n$ is a Gaussian noise. $T$ usually equals the number of time steps used for training, e.g., 1,000. 
DDPM sampling can be tedious, given the large number of sampling steps. Denoising diffusion implicit models' (DDIM) sampling method \cite{song2021denoising} significantly reduces the sampling steps. However, in our task, both DDPM and DDIM are prone to generate overly smooth samples at very low bitrates, i.e., 1 and 1.5 kbps, with some hallucination effects, such as missing or replaced phonemes. 

We believe the smoothing effect is caused by excessive noise reduction and insufficient assistance from the condition. Therefore, we propose a new sampling method, \textit{midway-infilling}, which improves the sampling quality and efficiency in two folds. 1) It starts sampling from a mid-point step $\tau<T$ rather than from $\bm x_T$, a random noise space, which reduces sampling steps by 10 to 20 times without sacrificing the sampling quality; 2) it implements a separate conditioning branch to impose stronger conditioning during sampling.

Midway-infilling is inspired by the infilling algorithm proposed in \cite{mittal2021symbolic}. Infilling aims to condition and steer the sampling steps on the unconditional diffusion model. 
In the original infilling process, an occluded sample $\bm s_0$ is provided. The diffusion process runs on $\bm s_0$ (infilling branch) to meet the time step of reversed sampling branch $\bm x_t$ (sampling branch), i.e., $\bm s_t = \mathcal{F}(\bm s_0, t)$. $\bm x_t$ and $\bm s_t$ are then interpolated with a certain ratio at each step.
Akin to infilling, the proposed midway-infilling has two branches. The difference is, instead of providing $\bm s_0$, we use condition $\bm h$ or its upsampled version to approximate $\bm s_{\tau}$, which is a midway variable of the infilling branch's Markov chain path. Accordingly, our infilling branches can run the reverse process from step $\tau$ to 0, as described in Algorithm \ref{alg:one}. 


\SetKwComment{Comment}{/* }{ */}

\begin{algorithm}[t]
\caption{Midway-Infilling}\label{alg:one}
\SetKwInput{KwInput}{Input}                
\SetKwInput{KwOutput}{Output} 
\KwInput{Conditioning vector $\bm h$, midway step $\tau$, interpolation ratio $\gamma$, sampling function $\mathcal{G}(\cdot)$}
$\bm s_{\tau} \gets \bm h, \quad \bm x_{\tau} \sim \mathcal{N}(0, 1), \quad \bm x_{\tau} = (1-\gamma) \bm x_{\tau}   + \gamma\bm s_{\tau}   $ \\
\For{$t = \tau \ldots 1$}{
    $\bm s_{t-1} = \mathcal{G}(\bm s_t, t, \bm h)$ -- Infilling branch\\
    $\bm x_{t-1} = \mathcal{G}(\bm x_t, t, \bm h)$ -- Sampling branch\\
    $\bm x_{t-1} = (1-\gamma)\bm x_{t-1}  + \gamma\bm s_{t-1} $ \\
  }
\Return $\bm x_0$ 
\end{algorithm}

\label{sec:format}

\section{Experiment Setup}
\label{sec:pagestyle}
\begin{figure*}[t]
     \centering
     \begin{subfigure}[b]{0.195\textwidth}
          \includegraphics[width=\textwidth]{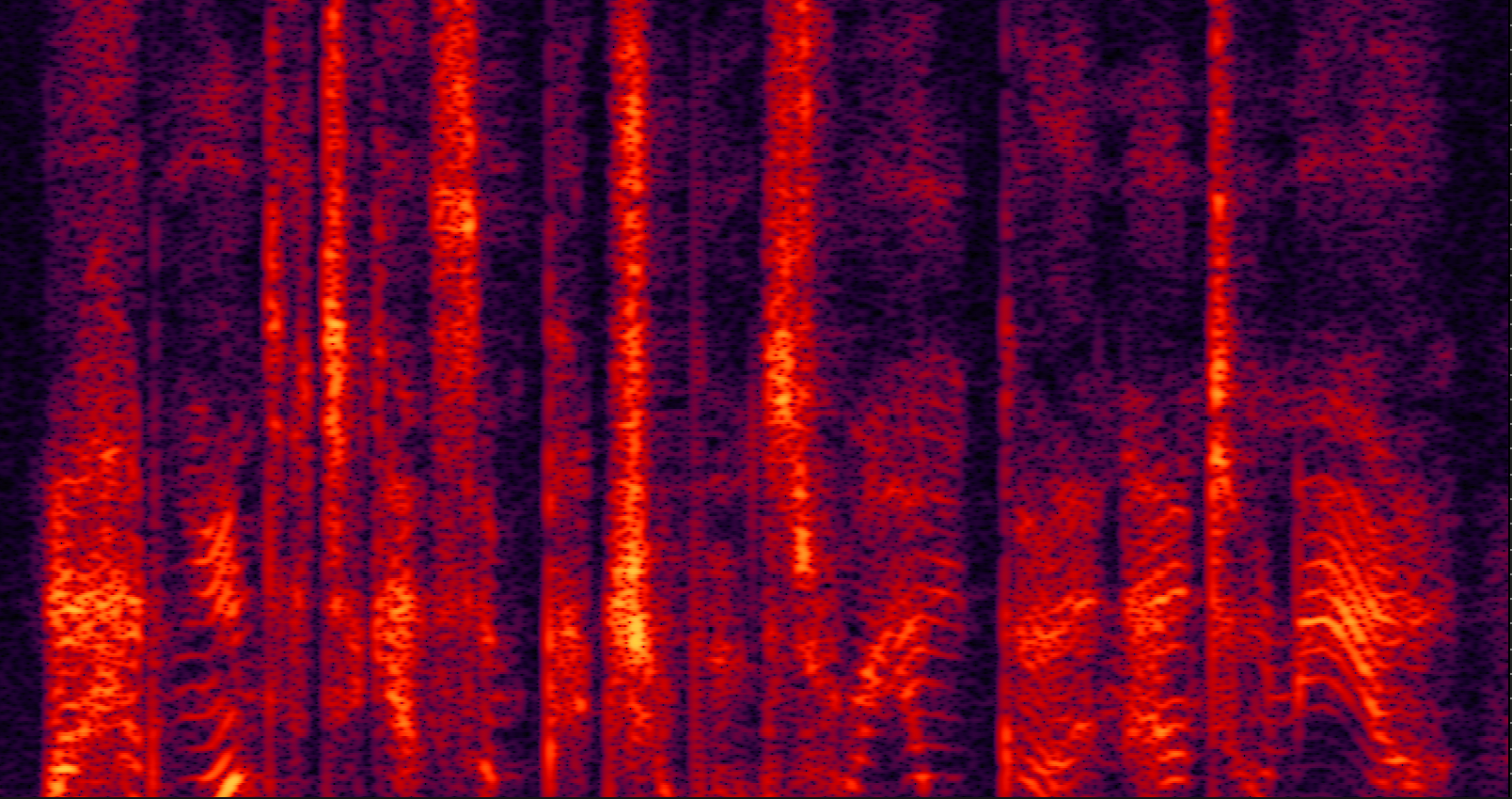}
         \caption{Reference}
         \label{fig:spec-ref}
     \end{subfigure}
     \hfill
     \begin{subfigure}[b]{0.195\textwidth}
          \includegraphics[width=\textwidth]{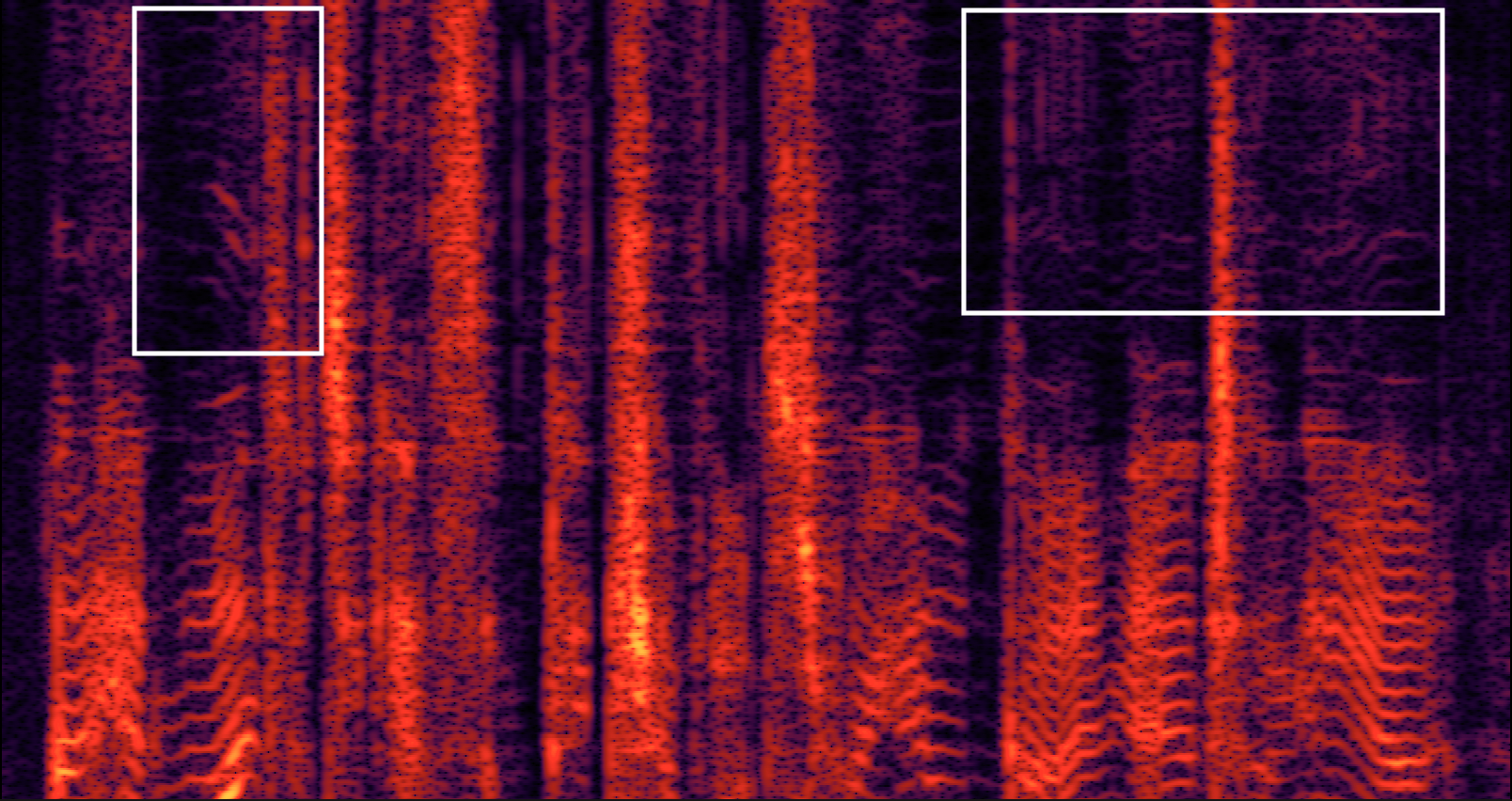}
         \caption{EnCodec@1.5kbps}
         \label{fig:spec-E1.5}
     \end{subfigure}
     \hfill
     \begin{subfigure}[b]{0.195\textwidth}
          \includegraphics[width=\textwidth]{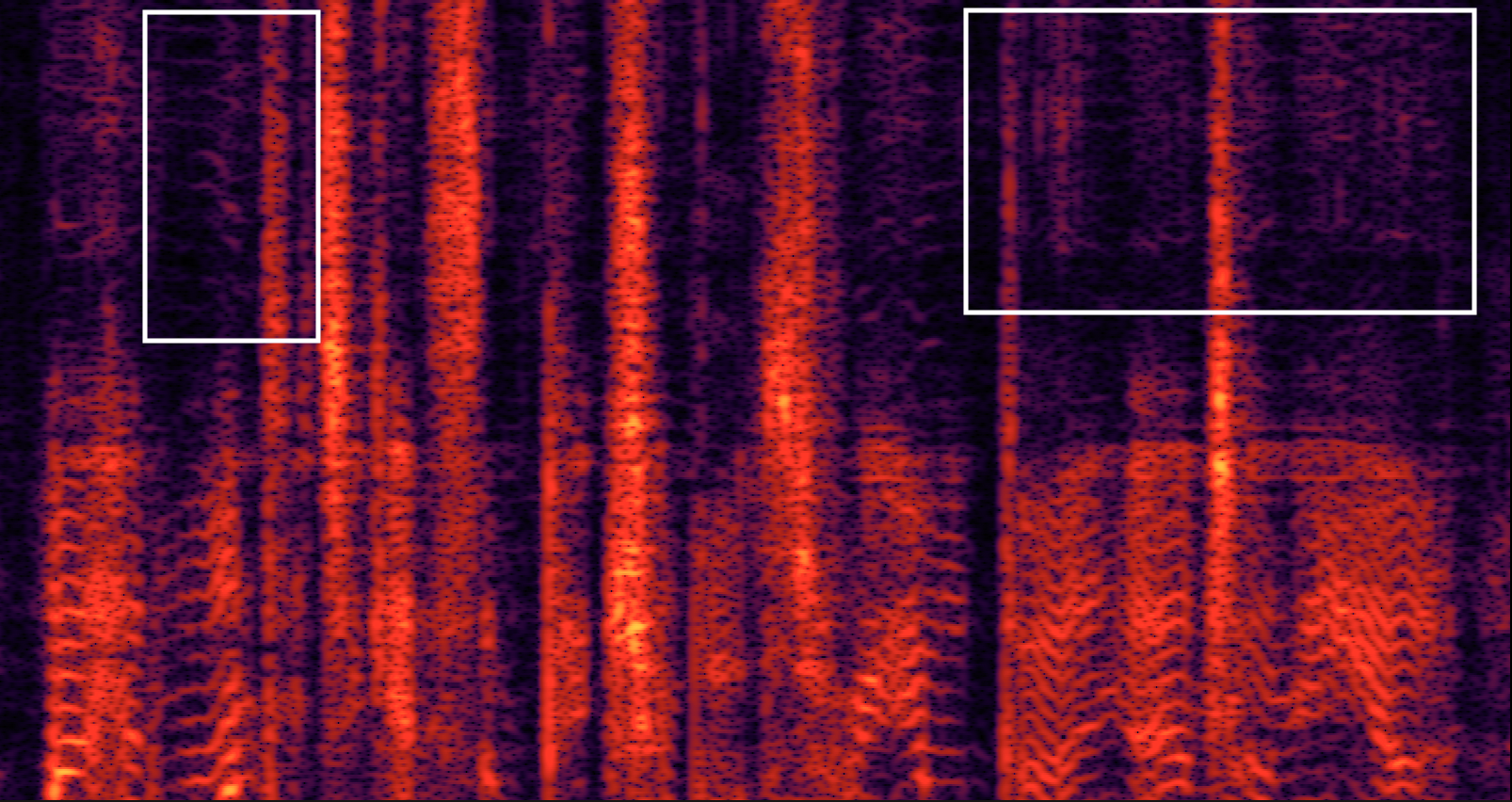}
         \caption{EnCodec@3kbps}
         \label{fig:spec-E3}
     \end{subfigure}
     \hfill
     \begin{subfigure}[b]{0.195\textwidth}
          \includegraphics[width=\textwidth]{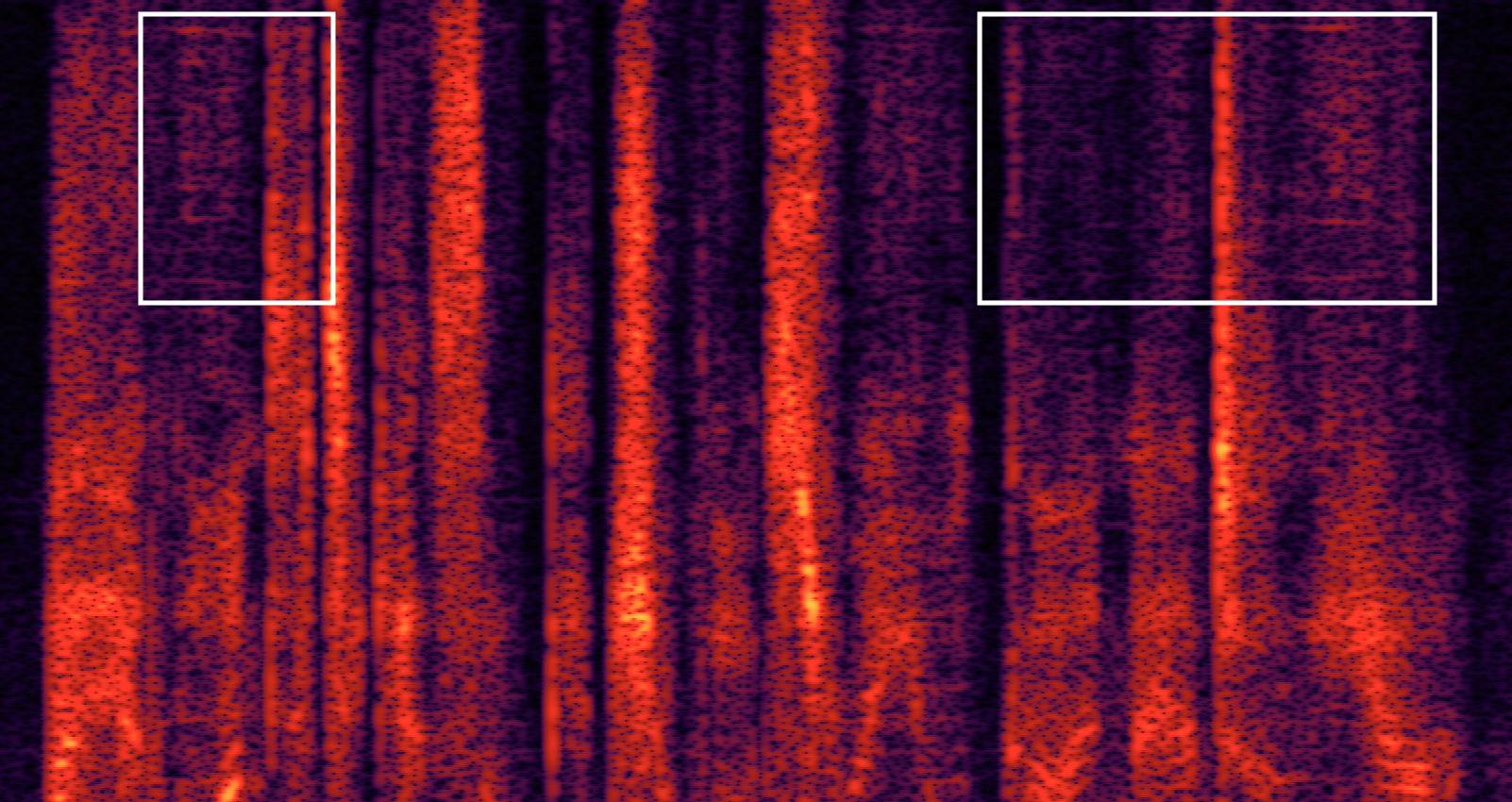}
         \caption{LaDiffCodec@1.5kbps}
         \label{fig:spec-D1.5}
     \end{subfigure}
     \hfill
     \begin{subfigure}[b]{0.195\textwidth}
          \includegraphics[width=\textwidth]{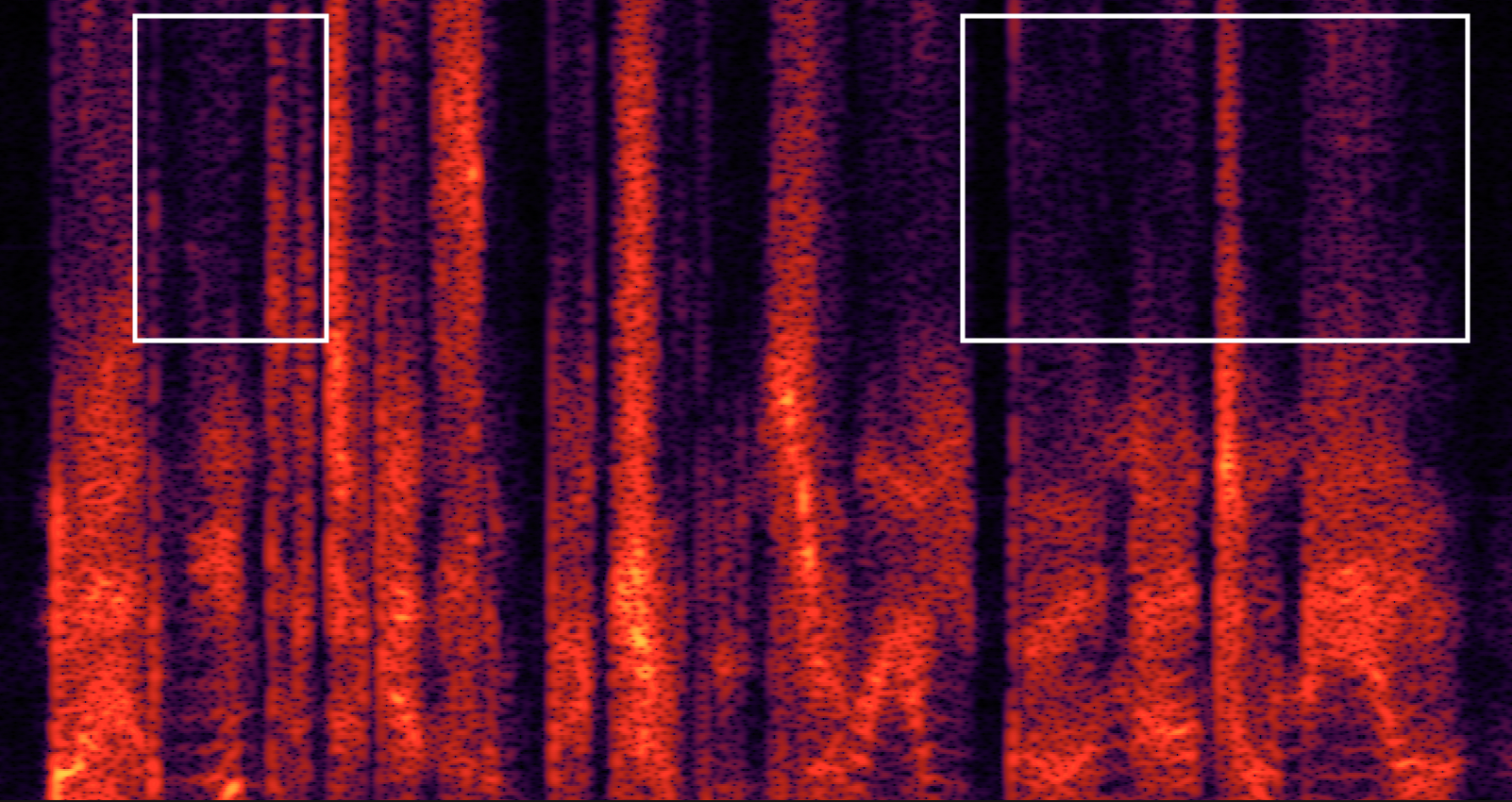}
         \caption{LaDiffCodec@3kbps}
         \label{fig:spec-D3}
     \end{subfigure}
    \caption{Spectrograms of the reference speech, and the coded versions by EnCodec and LaDiffcodec at 1.5kbps and 3kbps. The audio samples of these spectral are available at the sample page. White blocks point out the example areas where EnCodec shows aliasing artifacts.}
    \label{fig:spec}
\end{figure*}

\subsection{Model Design and Hyperparameter Setup}

\begin{figure}\label{fig:MUSHRA}
    \includegraphics[width=\columnwidth]{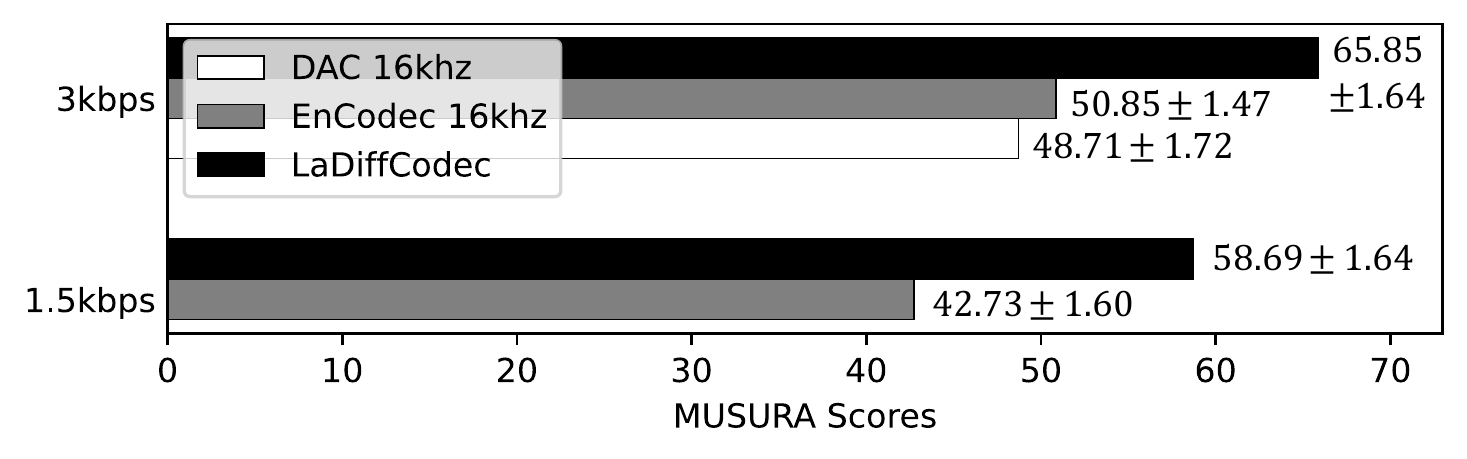}
    \caption{Average MUSHRA scores and their confident intervals of LaDiffCodec and baseline codec systems at 1.5kbps and 3kbps.}
    \vspace{-4mm}
    \label{fig:mushra}
\end{figure}

\begin{figure}[t]
     \centering
    \begin{subfigure}[b]{0.28\columnwidth}
          \includegraphics[height=.89in]{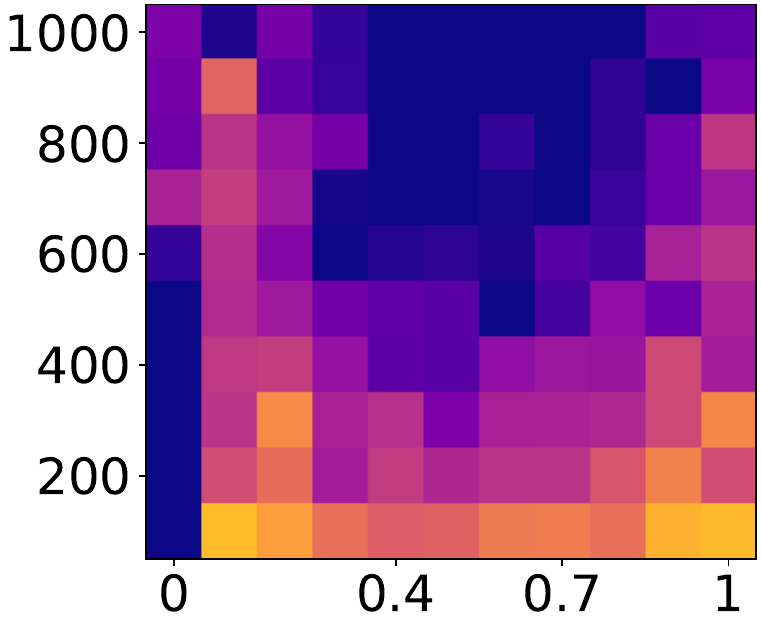}
         \caption{1kbps}
    \end{subfigure}
    \hfill
    \begin{subfigure}[b]{0.34\columnwidth}
          \includegraphics[height=0.92in]{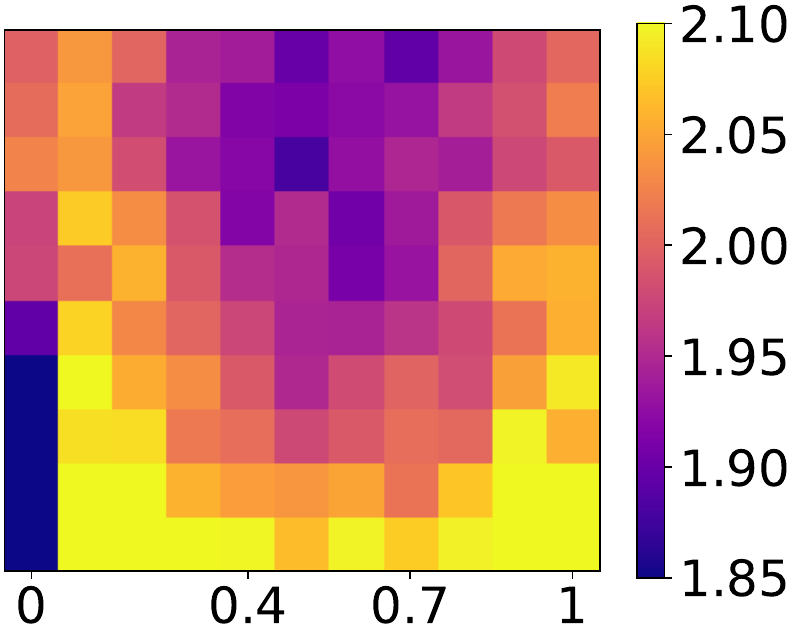}
         \caption{1.5kbps}
     \end{subfigure}
     \hfill
    \begin{subfigure}[b]{0.335\columnwidth}
          \includegraphics[height=.9in]{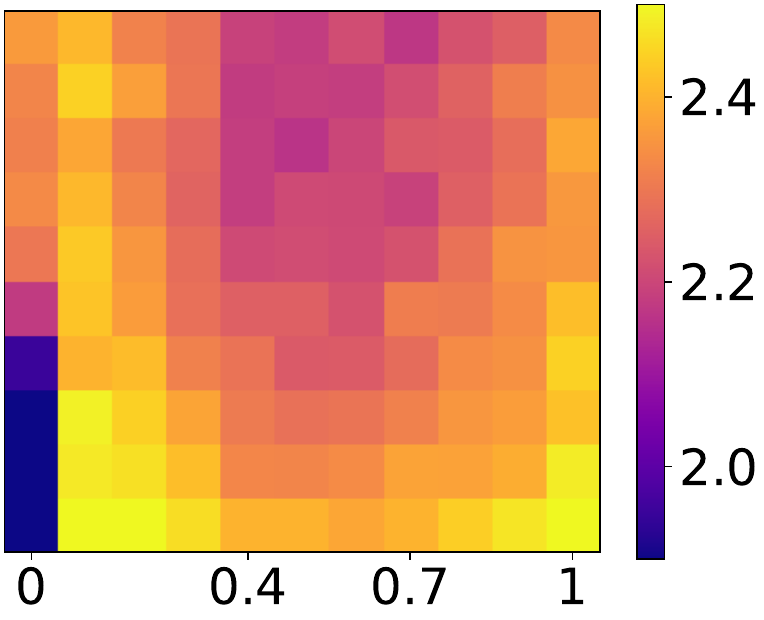}
         \caption{3kbps}
     \end{subfigure}
    \caption{PESQ scores of LaDiffCodec using different midway-infilling hyperparameters. X-axis denotes the mask ratios $\gamma$ from 0 to 1. Y-axis denotes the midway timestep $\tau$.}
    \label{fig:midway-sampling}
     \vspace{-4mm}
\end{figure}
In the forward process, we use $T\!=\!1000$ steps and set noise schedule linearly from $\beta_1\!=\!0.0001$ to $\beta_T\!=\!0.02$.
The reversed diffusion process is parameterized with a U-Net-based backbone, similar to \cite{rombach2022high, liu2023audioldm}. The U-Net is built with ResNet blocks, each containing three convolution layers and one four-head self-attention layer. The model comprises five encoder blocks, one middle block, and five decoder blocks. The channel dimensions of encoder blocks are $[128, 256, 256, 512, 512]$.
Decoder blocks have the reversed dimensions of encoder blocks, and the channel dimension of the middle block is 512. 
While AudioLDM \cite{liu2023audioldm} uses FiLM conditioning, we condition quantized tokens to the diffusion model by stacking the tokens $\bm h$ and model input $\bm z_t$ together, as we found it more effective than FiLM in this task. 
When the dimensionality of the continuous and discrete code spaces do not match, i.e., $D<L$, $\bm h$ is firstly upsampled with transposed convolutional layers. We scale each frame of the upsampled tokens to $[-1, 1]$ before conditioning them on the diffusion.

We use the proposed midway-infilling method for sampling, where the hyperparameters are shared in all bitrate cases to keep the usage simple. We set the midway step $\tau=100$ and $\gamma=0.3$. Once latent diffusion sampling is finished, the acquired continuous sample $\bm z_0$ is fed to the decoder, which maps it to the time-domain signal. Since $\tau=100$ is small, it takes $\sim$5.65 seconds to generate a 3.2-second sample. 

We retrain the non-streamable version of EnCodec with the 16 kHz Librispeech dataset \cite{PanayotovV2015Librispeech} as the discrete autoencoder. Its encoder and decoder have SEANet \cite{tagliasacchi2020seanet} as the backbone. The encoder downsamples input through four convolution layers with strides of size 2, 4, 5, and 8, respectively.  Using transposed convolution, the decoder upsamples the latent space in the reverse order.  In addition, we build a continuous autoencoder akin to EnCodec. We keep only one downsampling layer with a stride size of 8 to achieve higher dimension and, consequently, expressiveness.

\subsection{Data and training}
All experiments are conducted on the Libirspeech dataset, \texttt{train-\\clean-100} fold for training, and \texttt{dev-clean} for testing. Model training runs on the 3.2s sequences. The three components, i.e., discrete autoencoder (16khz EnCodec), continuous autoencoder, and the latent diffusion model, are trained separately. When training the latent diffusion model, both autoencoders are frozen. 
Our diffusion models are bitrate-specific. For example, LaDiffCodec at 1.5 kbps uses EnCodec's 1.5 kbps tokens as its condition.
We use Adam optimizer for all the training tasks, with a batch size of 20 and a learning rate 5e-5. It takes 6 hours to converge the autoencoder training and three days for the latent diffusion model on one NVIDIA A100 GPU.

\subsection{Evaluation and Ablation study}

Our MUSHRA-like subjective test \cite{mushra} compares LaDiffCodec with 16 kHz EnCodec at two bitrates, 1.5kbps and 3kbps. Sequences from a 16 kHz DAC at 3kbps are also included for comparison. 13 audio experts participated in and rated ten gender-balanced samples of 3 seconds.

All the ablation studies are evaluated with PESQ \cite{RixA2001pesq} on randomly picked 50 samples from the test set. We notice that PESQ doesn't reflect the real perceptual preference for different coding systems. However, as for the comparison of sequences generated from the same kind of system, they tend to exhibit a steady trend.

\section{Experimental Results}

\subsection{Comparison with other codec}

Figure \ref{fig:mushra} shows MUSHRA scores of different codec systems. We see LaDiffCodec surpasses EnCodec and DAC at both bitrates. Particularly, LaDiffCodec's 1.5 kbps samples are preferred by the subjects to the 3 kbps samples of the other codecs.

We believe that LaDiffCodec has two main traits that contribute to its superior performance. Firstly, it recovers coding artifacts. Lossy compression can cause various speech alterations and degradation. We observe that at the 1.5kbps, EnCodec starts to lose intelligibility, because some phonemes are not recovered precisely. DAC exhibits a similar artifact at 1.5 kbps (using three quantizer cookbooks). At 3kbps, while the intelligibility is better preserved, the baseline codecs still produce artifacts such as subtle background noise (DAC) or metallic and hissing sound (EnCodec).
LaDiffCodec can fix severe quantization artifacts by generating variables in the continuous latent space.  
The fact that the LD process works in the pre-learned latent distribution narrows its synthesis process to a more straightforward problem. In addition, the well-trained continuous latent space leads to higher speech reconstruction quality, eliminating the non-speech artifacts and distortion. 

Secondly, it generates more natural-sounding speech. Fig. \ref{fig:spec} presents spectrograms of the reference and various decoded versions. These low-bitrate EnCodec results suffer from losing high-frequency information due to the reduced expression space. Waveform coding, in particular, often experiences aliasing artifacts, as shown in \ref{fig:spec-E1.5} and \ref{fig:spec-E3}. In these spectrograms, the high-frequency area over $\sim$4 kHz shows a mirrored reflection of the lower frequency harmonics. These high-frequency aliasing effects can add unnatural artifacts to the reconstruction. In contrast, LaDiffCodec makes up some high-frequency energy and eludes the aliasing effect. As a result, it produces a more natural and pleasant sound. 
As we use DAC's public 16khz checkpoint, which is not re-trained on the Librispeech dataset, it appears less performed than the re-trained EnCodec.


\subsection{Ablation Studies}

\noindent\textbf{Hyperparameters of Midway-Infilling}:
This ablation explores different settings of midway-infilling hyperparameters. Smaller $\gamma$ correlates to less involvement of the condition branch $[\bm s_{\tau}, ..., \bm s_0]$ during sampling, while a large $\tau$ means the sampling process conducts more noise reduction. When $\gamma=0$ and $\tau=1000$, it is equivalent to DDPM's original sampling method. With the correct set of hyperparameters, midway-infilling gains higher PESQ than the original DDPM sampling. The leftmost columns of each graph present the degrading performance by reducing DDPM sampling steps, with no extra infilling branch involved (i.e., $\gamma=0$). The rightmost columns, on the other hand, present the infilling branch's sole contribution to sampling. According to the PESQ score, the best quality is obtained when sampling step $\tau$ is small and $\gamma$ is close to 0 or 1. 
Our perceptual rating aligns with PESQ concerning $\tau$. However, a large interpolation ratio $\gamma$ leads to a better phoneme-level reconstruction at the cost of less naturalness. A sequence of speech samples regarding different interpolation ratios can be found on our webpage.

\noindent\textbf{Latent Dimensionality}: Table \ref{tab:abla2} presents the PESQ scores of LaDiffCodec with different latent space dimensions. The total downsampling rate is the product of stride sizes. The first row shows results from the ordinary (non-latent) diffusion model, which samples in the time domain with no continuous autoencoder involved. All the experiments are made to run for the same amount of time. In comparison, the LD models outperform the time-domain diffusion method (stride$=1$), indicating that a reduced dimension and auxiliary feature learning can facilitate diffusion modeling, especially in this speech coding task. 
However, reduced dimension does not always lead to superior performance. When more downsampling layers are added to the continuous autoencoder, the latent space starts losing expression power, or becomes hard to model with the diffusion process.

\begin{table}[]
    \centering
    \small
    \tiny
    \resizebox{\columnwidth}{!}{%
    \begin{tabular}{c||c|c|c}
        Strides & @1kbps & @1.5kbps & @3kbps \\

        \hline
        [1] & 1.18 $\pm$ 0.04 & 1.20 $\pm$ 0.04 & 1.77 $\pm$ 0.19  \\
        \hline
        \textbf{[8]} & \textbf{1.81 $\pm$ 0.15} & 1.95 $\pm$ 0.15 & \textbf{2.23 $\pm$ 0.17} \\
        \hline
        [4, 8]& 1.71 $\pm$ 0.71 & \textbf{2.19 $\pm$  0.75 }& 2.16 $\pm$ 0.69 \\
        \hline
        [4, 5, 8] & 1.66 $\pm$ 0.11  & 1.71 $\pm$ 0.12  & 1.84 $\pm$ 0.10 \\
        \hline
        [2, 4, 5, 8] &  1.49 $\pm$ 0.09  & 1.65 $\pm$ 0.13  & 1.71 $\pm$ 0.12 \\
    \end{tabular}
    }
    \caption{Performance of diffusion models with different latent space dimension. The arrays in the first column present the stride sizes of each down-sampling layer in the continuous encoder.}
    \label{tab:abla2}
    \vspace{-3mm}
\end{table}

\section{Conclusion}
This work proposed LaDiffCodec and demonstrated the effectiveness of integrating waveform coding-based feature learning and latent diffusion model for high-quality, low-bitrate speech coding. By mapping the low-dimensional discrete speech token into high-dimensional continuous space using latent diffusion, the codec released the burden of upsampling and de-quantization from the decoder. It achieved improved speech quality with reduced artifact and increased naturalness. 
While mainly focusing on the low-bitrate scenarios, our work potentially sheds light on the high-fidelity codec-based generation. Our models provides a solution that enables using fewer codebooks for categorical generation, which reduces the task's difficulties without sacrificing output sound quality. 




\bibliographystyle{IEEEbib}
\bibliography{mjkim}

\end{document}